\documentclass[pra,
                twocolumn,
                preprintnumbers,
                amsmath,
                amssymb,
                showpacs,
                superscriptaddress]{revtex4-1}
\usepackage{graphicx} 
\usepackage{dcolumn}  
\usepackage{bm}       
\usepackage{amsfonts}
\usepackage[colorlinks=true,linkcolor=blue,citecolor=red,urlcolor=blue]{hyperref}
\usepackage{color}
\newcommand{\eq}[1]{Eq.~(\ref{#1})}
\newcommand{\ful}{\mbox{C$_{\mbox{\scriptsize{60}}}$}}
\newcommand{\full}{\mbox{C$_{\mbox{\scriptsize{240}}}$}}

\begin{document}

\title{Molecular size effects on diffraction resonances in positronium formation from fullerenes}

\author{Paul-Antoine Hervieux}
\email[]{hervieux@unistra.fr}
\affiliation{%
Universit\'e de Strasbourg, CNRS, Institut de Physique et Chimie des Mat\'eriaux de Strasbourg, 67000 Strasbourg, France}

\author{Anzumaan R. Chakraborty}
\thanks{Present address: Department of Physics, Missouri University of Science and Technology, Rolla, Missouri 65409, USA}
\affiliation{%
Department of Natural Sciences, D.L.\ Hubbard Center for Innovation,
Northwest Missouri State University, Maryville, Missouri 64468, USA}

\author{Himadri S. Chakraborty}
\email[]{himadri@nwmissouri.edu}
\affiliation{%
Department of Natural Sciences, D.L.\ Hubbard Center for Innovation,
Northwest Missouri State University, Maryville, Missouri 64468, USA}

\date{\today}



\begin{abstract}
We previously predicted [P.A.\ Hervieux et al.\ \pra\, \textbf{95}, 020701 (2017)] that owing to predominant electron capture by incoming positrons from the molecular shell, $\ful$ acts like a spherical diffractor inducing resonances in the positronium (Ps) formation as a function of the positron impact energy. By extending the study for a larger $\full$ fullerene target, we now demonstrate that the diffraction resonances compactify in energy in analogy with the shrinking fringe separation for larger slit size in classical single-slit experiment. The result brings further impetus for conducting Ps spectroscopic experiments with fullerene targets, including target- and/or captured-level differential measurements. The ground states of the fullerenes are modeled in a spherical jellium frame of the local density approximation (LDA) method with the exchange-correlation functional based on the van Leeuween and Baerends (LB94) model potential, while the positron impact and Ps formation are treated in the continuum distorted-wave final state (CDW-FS) approximation.
\end{abstract}

\maketitle

\section{Introduction}

Formation of an exotic and quasi-stable electron-positron bound-pair, positronium (Ps) -- a pure leptonic atom, by shooting positrons at matter is a fundamental process in nature. The Ps formation channel covers a large portion of the positron scattering cross section from simple atoms and molecules~\cite{laricchia2008}, while exhibiting even higher success rates on thin films and surfaces~\cite{schultz1988}. Besides probing material structure and reaction mechanism, applied interests in the Ps formation are aplenty. Ps perishes \textit{via} a unique electron-positron annihilation pathway~\cite{green2015,kauppila2004} with astrophysical~\cite{higdon2009,prantzos2011}, materials~\cite{kavetskyy2014}, and pharmaceutical~\cite{dong2015} interests. Efficient Ps formation is the precursor of the production of dipositronium molecules~\cite{cassidy2007} and antihydrogen atoms~\cite{ferragut2010, mcconnell2016} required to study the effect of gravity on antimatter~\cite{crivelli2014, Perez2015}. Possible production of Bose-Einstein condensate of Ps has also been predicted~\cite{shu2016,mor2014}, besides the importance of Ps in diagnosing porous materials~\cite{cassidy2008} as well as in probing bound-state QED effects~\cite{karshenboim2004}. Furthermore, similarities in the formalism for Ps formation in matter with the exciton theory in quantum dots have recently been shown~\cite{pietrow2018}.

There is an abundance of theoretical investigations in the literature to calculate Ps formation from a wide varieties of target. This includes atomic targets: (i) the hydrogen atom using variants of coupled-channel methods~\cite{yamanaka2001,kamali2001} and multichannel Schwinger's principle~\cite{kar2000}; (ii) noble gas atoms in the distorted-wave~\cite{sen2009}, the boundary-corrected Born~\cite{ghanbari2013}, and the relativistic optical potential method~\cite{mceachran2013}; and (iii) alkali metal atoms in the optical potential approach~\cite{gianturco1996} and in the classical trajectory Monte Carlo method when the targets are in Debye plasma environments~\cite{pandey2016}. Among these studies, ab\,initio close-coupling calculations, pioneered by Walters and collaborators~\cite{walters1996}, have in general been very successful~\cite{kadyrov2016}. Relatively limited calculations with molecular targets include: (i) the molecular hydrogen by utilizing the convergent close-coupling theory~\cite{kadyrov2016} and a model coupled-channel formalism~\cite{biswas2002}; and (ii) the water molecule in the continuum distorted-wave final-state approximation~\cite{hervieux2006}.

Precision experimental techniques to measure Ps formation signals have also been achieved by impinging positrons into varieties of materials, such as, atomic and molecular gases~\cite{garner1998,machacek2016}, polyatomic molecules~\cite{sueoka2000}, molecular solids~\cite{eldrop1983}, liquids and polymers~\cite{wang1998}, zeolites~\cite{cabral-prieto2013}, metal surfaces and films~\cite{cooper2016,jones2016}, metal-organic-frameworks~\cite{crivelli2014mof,jones2015}, and embedded mesostructures~\cite{andersen2016}. To facilitate precision measurements of gravitational free fall of antimatter as well as the optical spectrum of Ps, Doppler-corrected Balmer spectroscopy of Rydberg Ps has been applied~\cite{jones2014}. Recently high yields of laser assisted production of low-energy excited Ps is achieved in the interaction of cold-trapped positrons with Rydberg excited Cs atom~\cite{mcconnell2016}.

In spite of such broad landscape of Ps research, studies of Ps formation by implanting positrons in vapor or solid phase nanoparticles are rather scarce. On the other hand,  clusters and nanostructures straddle the boundary between atoms and condensed matters which enable them to exhibit hybridized properties of both domains often revealing remarkable behaviors with unusual spectroscopy~\cite{jaenkaelae2011}. A lonely theoretical study of Ps formation using the Na cluster targets was made about two decades ago~\cite{fojon2001}. Recently, however, pilot studies of Ps formation from the $\ful$ fullerene has been published by us~\cite{hervieux2017,chakraborty2017}. It has been shown that the formation of a gas of delocalized electrons within a finite nanoscopic region of more defined short-range boundary at the $\ful$ shell, in contrast of a long-range, diffused Coulombic decay of atomic and molecular electron densities, ensures predominant electron capture from local regions in space. This leads to diffraction in the capture amplitude, particularly at positron energies that cannot excite plasmon modes. Indeed, Ref.\,[\onlinecite{hervieux2017}] revealed a series of diffraction resonances in Ps formation from $\ful$ that may be observed in the experiment both in ground and excited state Ps formation.

In general, the Ps formation from fullerenes can be singularly attractive due to fullerene's eminent symmetry and stability in room-temperature, and its previous track record of success in photo-spectroscopic experiments~\cite{ruedel2002}. In this communication, we extend our study of the Ps formation to a larger fullerene $\full$. Ps($1s$) formation for captures from various $\full$ molecular orbitals also show diffraction resonances as a function of positron impact energy. What is particularly noticeable going from $\ful$ to $\full$ is an appropriate reduction of the energy separation between the resonances due to the increase of the molecular size as a strong signature of the underlying diffraction process. In fact, Fourier transforms of the resonant signals expressed in the target recoil momentum scale map the fullerene radii very well. The following section presents methods applied to carry out the numerical calculations. Section III presents and discusses the main results, while the final section concludes the article with some words to encourage future experiments.

\section{Description of the methods}

\subsection{LDA to model fullerene ground states}

The details of the method follow the framework as described in Ref.\,[\onlinecite{choi2017}]. The jellium potentials, $V_{\mbox{\scriptsize jel}}(\vec{r})$, representing 60 and 240 C$^{4+}$ ions, respectively for $\ful$ and $\full$, are constructed by smearing the total positive charge over spherical shells with radius $r_c$ and thickness $\Delta$. $r_c$ is taken to be the known radius of each molecule: 3.54 \AA\, (6.7 a.u.) for $\ful$~\cite{ruedel2002} and 7.14 \AA\, (13.5 a.u.) for $\full$~\cite{Lu94}. A constant pseudopotential $V_0$ is added to the jellium for quantitative accuracy~\cite{puska93}. The Kohn-Sham equations for systems of 240 and 960 electrons, made up of four valence ($2s^22p^2$) electrons from each carbon atom, are then solved to obtain the single electron ground state orbitals in the local density approximation (LDA). The parameters $V_0$ and $\Delta$ are determined by requiring both charge neutrality and obtaining the experimental value~\cite{devries1992} (for $\ful$) and the known theoretical value~\cite{white93} (for $\full$) of the first ionization thresholds. Consequently, the values of $\Delta$ are found to be 1.30 \AA\, and 1.50 \AA\, respectively for $\ful$ and $\full$.

Using the single-particle density $\rho(\vec{r})$ the LDA potential can be written as,
\begin{equation}\label{lda-pot}
V_{\scriptsize \mbox{LDA}}(\vec{r}) = V_{\mbox{\scriptsize jel}}(\vec{r}) + \int d\vec{r}'\frac{\rho(\vec{r}')}{|\vec{r}-\vec{r}'|} + V_{\scriptsize \mbox{XC}}[\rho(\vec{r})],
\end{equation}
where the 2nd and 3rd terms on the right are the direct Hartree and the basic exchange-correlation (xc) components. This basic xc functional $V_{\scriptsize \mbox{XC}}$ is parametrized directly from $\rho(\vec{r})$ by the following formula~\cite{gunnarsson76}:
\begin{eqnarray}\label{gl}
V_{\scriptsize \mbox{XC}}[\rho(\vec{r})] & = & -\left(\frac{3\rho(\vec{r})}{\pi}\right)^{1/3} \nonumber \\
 &-& 0.0333\log\left[1 + 11.4\left(\frac{4\pi\rho(\vec{r})}{3}\right)^{1/3}\right],
\end{eqnarray}
in which the first term on the right is exactly derivable by a variational approach from the HF exchange energy of a uniform electron system with a uniform positively charged background and the second term is the so called correlation potential, a quantity not borne in HF formalism. The xc-functional that utilizes \eq{gl} is then further refined by adding a parametrized potential~\cite{van1994exchange} in terms of the reduced density and its gradient $\nabla \rho$ as follows,
\begin{equation}\label{lb94}
V_{\mbox{\scriptsize LB}} = -\beta{[\rho(\vec{r})]^{1/3}}\frac{(\xi X)^2}{1+3\beta\xi X \sinh^{-1}(\xi X)},
\end{equation}
where $\beta = 0.05$ is empirical and $X=[\nabla \rho]/\rho^{4/3}$. The parameter $\xi$ is a factor arising in transition from the spin-polarized to spin-unpolarized form~\cite{oliver79}. This method of gradient-correction to the xc-functional, termed as LB94, is more built into the theory and leads to a considerable improvement in the asymptotic behavior of the electron by comparing well with the exact Kohn-Sham potentials calculated from correlated densities.
\begin{figure}[h!]
\vskip 0.7cm
\includegraphics[width=8.5cm]{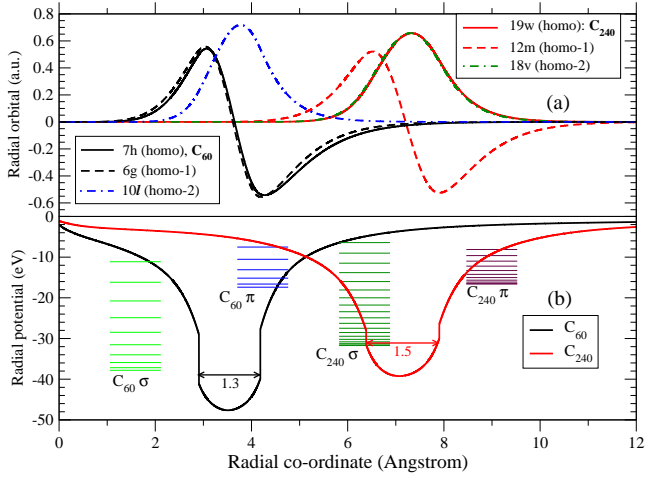}
\caption{(Color online) (a) Ground state radial wavefunctions for homo, homo-1, and homo-2 of $\ful$ and $\full$ calculated in LDA. (b) Corresponding radial potentials are shown and shell widths are identified. Energy bands of $\sigma$ and $\pi$ characters (see text) are illustrated}
\label{figure1}
\end{figure}

We show the ground state radial potentials and bands of $\ful$ and $\full$ in Fig.\,1(b). $\ful$ produced bands of six $\pi$ (one radial node) and ten $\sigma$ (nodeless) states. Among these the homo and \mbox{homo-1} levels are of 7$h$ ($\ell=5$) and 6$g$ ($\ell=4$) $\pi$ character respectively -- a result known from the quantum chemical calculations~\cite{troullier1992} supported by direct and inverse photoemission spectra~\cite{weaver1991}, and from energy-resolved electron-momentum density measurements~\cite{vos1997}. The \mbox{homo-2} level is of 10$l$ ($\ell=9$) $\sigma$ character. The LDA radial wavefunctions for these three outer states are shown in Fig.\,1(a). Linear response type calculations using this ground state basis well explained measured photoemission response of $\ful$ at the plasmon excitation energies~\cite{choi2017,scully2005}. Similar calculations at higher energies also supported an effective fullerene width accessed in the photoemission experiment~\cite{ruedel2002}. These general ground state properties also hold good for $\full$ that produced bands of nineteen $\sigma$ and eleven $\pi$ states, while its homo and \mbox{homo-2} are of 19$w$ ($\ell=18$) and 18$v$ ($\ell=17$) $\sigma$ character respectively with \mbox{homo-1} being 12$m$ ($\ell=10$) $\pi$ (Fig.\,1).

\subsection{CDW-FS to model Ps formation}

We consider an incoming positron of momentum $\vec{k}_i$ which captures an electron from a C$_N$ fullerene bound state $\phi_i(\vec{r}_-)$ to form a Ps state $\phi_f(\vec{\rho})$. As illustrated in Fig.\,2, the positron and electron position vectors, $\vec{r}_+$ and  $\vec{r}_-$ respectively, originate from the center of the C$_N^+$ ion so that $\vec{\rho}=\vec{r}_+ - \vec{r}_-$ is their relative position vector. $\vec{k}_{+(-)}$ denote positron (electron) outgoing momenta in Ps that are equal, resulting $\vec{k}_{\beta} = 2\vec{k}_{+(-)}$ to be the momentum of Ps itself. To be exact, all the momenta are reduced momenta. Since we access energies above fullerene plasmon resonances, the many-body effect is not important, justifying the use of mean-field LDA wavefunctions and potentials for C$_N$ (subsection IIA) in the framework of independent particle model. In that frame, the {\em prior} form of the Ps formation amplitude can be given in the continuum distorted-wave final-state (CDW-FS) approximation~\cite{fojon1996,fojon2001} as,
\begin{equation}\label{tot-amp}
T^-_{\alpha \beta}(\vec{k}_i) \sim \int d\vec{r}_- F_{\vec{k}_-}^{(-)^\ast}(\vec{r}_-) W(\vec{r}_-;\vec{k}_i) \phi_i(\vec{r}_-),
\end{equation}
in which
\begin{eqnarray}\label{pos-amp}
W(\vec{r}_-;\vec{k}_i) &=& \int d\vec{r}_+ F_{\vec{k}_+}^{(-)^\ast}(\vec{r}_+) \phi_f^\ast(\vec{\rho})\nonumber \\
                       && \times \left[V_i^{sc}(r_+) - \frac{1}{\rho}\right]F_{\vec{k}_i}^{(+)}(\vec{r}_+),
\end{eqnarray}
with,
\begin{subequations}\label{dist}
\begin{eqnarray}\label{dist1}
F_{\vec{k}_i}^{(+)} &=& N_{\nu _{\alpha}^{\prime }}^{+}\;\exp (i{\vec{k}}_{i}\cdot {\vec{r}_{+}})\nonumber \\
&\times& _{1}F_{1}(-i\nu _{\alpha}^{\prime };1;-i{\vec{k}}_{i}\cdot {\vec{r}_{+}}+ik_{i}r_{+})
\end{eqnarray}
and
\begin{eqnarray}\label{dist2}
F_{\vec{k}_-}^{(-)}\;F_{\vec{k}_+}^{(-)} &=& N_{\beta _{-}}^{-}\;N_{\beta _{+}}^{-}\;\exp (i({\vec{k}}_{-} \cdot {\vec{r}}_{-} + \vec{k}_{+} \cdot {\vec{r}}_{+}))  \nonumber \\
&\times& _{1}F_{1}(-i\beta_{-};1;-i{\vec{k}}_{-} \cdot {\vec{r}}_{-}-ik_{-}r_{-}) \nonumber \\
&\times& _{1}F_{1}(i\beta _{+};1;-i{\vec{k}}_{+} \cdot {\vec{r}}_{+}-ik_{+}r_{+}).
\end{eqnarray}
\end{subequations}
\begin{figure}[h!]
\includegraphics[angle=0,width=8.5cm]{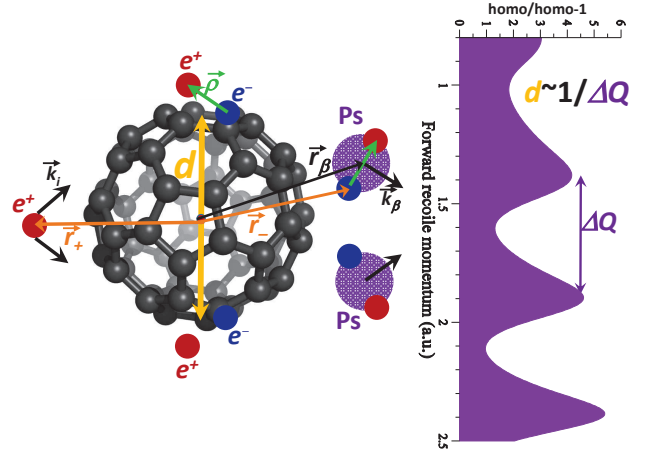}
	\caption{(Color online) A schematic diagram of the diffraction mechanism in the Ps formation from $\ful$. Position vectors of positron, electron, and Ps from the center of the molecular cation, and the electron-positron relative position vector $\vec{\rho}$ in Ps are schematically shown. The incoming and outgoing momentum vectors are indicated. Diffraction resonances in the ratio between Ps formation cross sections for $\ful$ homo and homo-1 captures are included.}
	\label{figure2}
\end{figure}

We have defined~\cite{fojon1996},
\begin{eqnarray}
\beta _{+} &\simeq &\beta _{-}=\frac{(Z+1)\mu _{\beta }}{k_{\beta }}=\frac{(Z+1)\mu _{\alpha }}{k_{\pm }}\; \\
\nu _{\alpha }^{\prime } &=&\frac{Z\mu _{\alpha }}{k_{i }},
\end{eqnarray}
where $Z$ is the net charge of the target (here $Z=0$), and the reduced masses are $\mu _{\alpha }\simeq 1$ and $\mu _{\beta }\simeq 2$, so that
\begin{eqnarray}
N_{\beta _{\pm }}^{-} &=&\Gamma (1\mp i\beta _{\pm })\;\exp (\mp \frac \pi 2%
\beta _{\pm })\; \\
N_{\nu _\alpha ^{\prime }}^{+} &=&\Gamma (1+i\nu _\alpha ^{\prime })\;\exp (-%
\frac \pi 2\nu _\alpha ^{\prime })\;.
\end{eqnarray}
The positron scattering potential in \eq{pos-amp} is given by,
\begin{equation}
V_i^{sc} = V_{i}^{sr}(r_{+}) + \frac{1}{r_{+}} \;,
\end{equation}
where $V_{i}^{sr}$ is the short-range part of the positron-residual target interaction associated to the fullerene orbital labeled $i$ so that
\begin{equation}
V_{i}^{sr}(r_{+})=-V_{\mbox{\scriptsize jel}}(r_{+})-\sum_{k=1;k\ne i}^{N_{\mathrm{\mathrm{orb}}}}V_{\mbox{\scriptsize H}%
}[\rho_{k}({\vec{r}})]-\frac{(Z+1)}{r_{+}}\;,
\end{equation}
in which $N_{\mathrm{\mathrm{orb}}}$ is the number of fullerene orbitals (see subsection IIA), and $V_{\mbox{H}}$ and $V_{\mbox{\scriptsize jel}}$ are respectively the Hartree and the jellium potential as in \eq{lda-pot}.

We, thus, write the {\em prior} version of the CDW-FS amplitude \eq{tot-amp} as~\cite{fojon1996,fojon1997}
\begin{widetext}
\begin{eqnarray}
T_{\alpha \beta }^{-} &=& N_{\nu _\alpha ^{\prime }}^{+}\;N_{\beta
_{+}}^{-*}\;N_{\beta _{-}}^{-*}\;\int d{\vec r}_{+}\;d{\vec r}_{-}\;\; \exp \left\{ i{%
\vec k}_i \cdot {\vec r}_{+} - i{\vec{k}}_{+} \cdot {\vec{r}}_{+} -i \vec{k}_{-} \cdot {\vec{r}}_{-} \right\}\;
_1F_1(-i\nu _\alpha ^{\prime };1;-i{\vec k}_i \cdot {\vec r}_{+} + k_i r_{+}) \; \phi _{i}({\vec{r}_{-}})
\nonumber \\
&\times& \left( V_{i}^{sr}(r_{+}) + \frac {1}{r_{+}}-\frac 1\rho \right)\;\phi_f^{*}(\vec{\rho})\;_{1}F_{1}(-i\beta_{+};1;i{\vec{k}}_{+} \cdot {\vec{r}}_{+}+ik_{+}r_{+})\;
_{1}F_{1}(i\beta_{-};1;i{\vec{k}}_{-} \cdot {\vec{r}}_{-}+ik_{-}r_{-}) \;.
\end{eqnarray}
\end{widetext}
In order to evaluate the amplitude, a partial wave expansion technique introduced in~\cite{fojon1996} has been employed.

The initial fullerene orbital is
\begin{equation}
\phi _{i}({\vec{r}}_{-})=R_{n_{t}\ell_{t}}(r_{-})\;Y_{\ell_{t},m_{t}}(\hat{r}_{-}),
\end{equation}
where $n_{t},\ell_{t}$ and $m_{t}$ are the quantum numbers. The final wavefunction is given by
\begin{equation}
\phi_{f}(\vec{\rho })=\frac{1}{\sqrt{2}}\exp (-\rho  /2)\text{ }Y_{0,0}(\hat{\rho })\equiv {\tilde{R}}_{1s}(\rho  )\ Y_{0,0}(\hat{\rho }),
\end{equation}
since the ground state, $1s$, of the Ps atom is considered in the present work. The angle differential cross section (DCS) for the capture then reads
\begin{equation}\label{dcs-nlm}
\left[ \frac{d\sigma }{d\Omega }\right] _{n_{t}\ell_{t}m_{t}}=\frac{1}{4\pi ^{2}%
}\;\frac{k_{\beta }}{k_{i }}\;\mu _{\alpha }\mu _{\beta }\;\left|
T_{\alpha \beta }^{-}\right| ^{2}
\end{equation}
with
\begin{eqnarray}
{\left| T_{\alpha \beta }^{-}\right| }^{2} &=& \frac{(4\pi )^{3}}{(k_{i}k_{+}k_{-})^{2}}{\hat{l}_{t}}\nonumber \\
&\times& \left|\sum_{l_{i}L}\;i^{l_{i}}\;e^{i\delta _{l_{i}}}{\hat{l}_{i}}{\hat{L}}%
^{1/2}(-1)^{L}\;{\cal S}_{l_{i}L}\;Y_{L,m_{t}}(\hat{k}_{\beta })\right| ^{2}
\end{eqnarray}
where the notation $\widehat{l}=2l+1$ has been used. Moreover, we have
defined
\begin{equation}
{\cal S}_{l_{i}L}=\sum_{ll^{\prime }l_{f}}\;i^{-l-l_{f}}\;e^{i(\delta
_{l}+\delta _{l_{f}})}\;(-1)^{l^{\prime }}\;{\cal A}_{l_{i}L}^{ll^{\prime
}l_{f}}\;{\cal R}_{l_{i}l_{f}}^{ll^{\prime }}
\end{equation}
with
\begin{eqnarray}
{\cal A}_{l_{i}L}^{ll^{\prime }l_{f}} &=& {\hat{l}}\;{\hat{l}^{\prime }}\;{\hat{l%
}_{f}}\;\left(
\begin{array}{ccc}
\ell_{t} & l & l^{\prime } \\
0 & 0 & 0
\end{array}
\right) \left(
\begin{array}{ccc}
l_{i} & l^{\prime } & l_{f} \\
0 & 0 & 0
\end{array}
\right) \left(
\begin{array}{ccc}
l & l_{f} & L \\
0 & 0 & 0
\end{array}
\right) \nonumber \\
&\times& \left(
\begin{array}{ccc}
l_{i} & L & \ell_{t} \\
0 & -m_{t} & m_{t}
\end{array}
\right) \left\{
\begin{array}{ccc}
l_{i} & L & \ell_{t} \\
l & l^{\prime } & l_{f}
\end{array}
\right\} \;,
\end{eqnarray}
\begin{equation}
{\cal R}_{l_{i}l_{f}}^{ll^{\prime }}=\int_{0}^{\infty }F_{l_{i}}(k_{i
}r_{+})\;{\cal V}_{ll^{\prime }}(r_{+})\;F_{l_{f}}(k_{+} r_{+})\;dr_{+}\;,  \label{rad}
\end{equation}
\begin{equation}
{\cal V}_{ll^{\prime }}(r_{+})=\int_{0}^{\infty
}r_{-}\;R_{n_{t}\ell_{t}}(r_{-})\;J_{l^{\prime }}(r_{-};r_{+})\;F_{l}(k_{-}r_{-})\;dr_{-}\;,
\label{tra}
\end{equation}
\begin{eqnarray}
J_{l^{\prime }}(r_{-};r_{+}) &=& \frac{1}{2}\;\int_{-1}^{+1}{\tilde{R}}_{1s}(\rho
)\;\left( V_{i}^{sr}(r_{+}) + \frac{1}{r_{+}} -\frac{1}{\rho} \right)\nonumber \\
&\times& P_{l^{\prime }}(u)\;du\;,  \label{Jl}
\end{eqnarray}
and
\begin{equation}
\rho ={\left( r_{-}^{2}+r_{+}^{2}-2r_{-}r_{+}u \right) }^{1/2}\;.
\end{equation}
The functions $F_{l}(k_{\pm }r)$ and $F_{l}(k_{i }r)$ are the Coulomb
radial wave functions with the Sommerfeld parameters $\eta =\beta _{\pm }$
and $\eta =\nu _{\alpha }^{\prime }$, respectively. The phase shifts $%
\delta _{l}$ are the usual Coulomb phase shifts $\delta _{l}=\arg \Gamma
(l+1+i\eta )$. $P_{l}$ indicates the Legendre polynomial of degree $l$.

Upon averaging \eq{dcs-nlm} over $m_t$ and denoting the electron occupancy number of the fullerene $(n_t \ell_t)$ state by $\mathrm{occ}(n_t \ell_t)$, we obtain
\begin{equation}\label{diff}
\left[ \frac{d\sigma }{d\Omega }\right] _{n_t \ell_t}=\frac{\mathrm{occ}(n_t \ell_t)}{2(2\ell_t+1)}
\times \sum_{m_{t}}\left[ \frac{d\sigma }{d\Omega }\right] _{n_{t}\ell_{t}m_{t}} \;.
\end{equation}
Finally, the angle-integrated cross section is evaluated as,
\begin{eqnarray}
\left[ \sigma \right] _{n_t \ell_t} &=& \int_{0}^{\pi} \sin(\theta) d\theta\int_{0}^{2\pi}d\varphi \left[ \frac{d\sigma }{d\Omega }\right] _{n_t \ell_t} \nonumber \\
&=& \mathrm{occ}(i) \times \frac{16\pi \mu _{\alpha }\mu
_{\beta }k_{\beta }}{k_{i }^{3}\left( k_{+}k_{-}\right) ^{2}}\;\sum_{l_{i}L}\;{\hat{l}_{i}}{\hat{L}}\;{\tilde{{\cal S}}}%
_{l_{i}L}{\tilde{{\cal S}}}_{l_{i}L}^{*}, \label{x-section}
\end{eqnarray}
where $(\theta, \varphi)$ are the angles of ${\vec k}_\beta$ (with respect to the incoming positron direction defined by $\vec{k}_{i}$ and which is considered to be along the $z$-axis), and
\begin{equation}
{\tilde{{\cal S}}}_{l_{i}L}=\sum_{ll^{\prime
}l_{f}}\;i^{-l-l_{f}}\;e^{i(\delta _{l}+\delta _{l_{f}})}\;(-1)^{l^{\prime
}}\;{\tilde{{\cal A}}}_{l_{i}L}^{ll^{\prime }l_{f}}\;{\cal R}%
_{l_{i}l_{f}}^{ll^{\prime }} \;,
\end{equation}
\begin{eqnarray}
{\tilde{{\cal A}}}_{l_{i}L}^{ll^{\prime }l_{f}} &=& {\hat{l}}\;{\hat{l}_{f}}%
\;\left(
\begin{array}{ccc}
\ell_{t} & l & l^{\prime } \\
0 & 0 & 0
\end{array}
\right) \left(
\begin{array}{ccc}
l_{i} & l^{\prime } & l_{f} \\
0 & 0 & 0
\end{array}
\right) \left(
\begin{array}{ccc}
l & l_{f} & L \\
0 & 0 & 0
\end{array}
\right) \nonumber \\
&\times& \left\{
\begin{array}{ccc}
l_{i} & L & \ell_{t} \\
l & l^{\prime } & l_{f}
\end{array}
\right\} \;.
\end{eqnarray}

\section{Results and discussions}

We compute Ps($1s$) formation cross sections, \eq{x-section}, for the capture from various fullerene levels as a function of the positron impact energy in LDA + CDW-FS scheme discussed above. Cross sections show trains of shape resonances. These resonances emerge from a diffraction mechanism based on the fullerene molecular structure. An elegant analytic interpretation of this diffraction effect is given in Ref.\,[\onlinecite{hervieux2017}] which we briefly review here. This treatment assumed plane waves, instead of three distorted Coulomb continuum waves in Eqs.\,(\ref{dist}). It also assumed Ps formations in the forward direction, which was found to be the most dominant direction in both earlier~\cite{tang1993} and contemporary experiments~\cite{shipman2015}. It was further noted: (i) Large values of $V_i^{sc}$ [Fig.\,5(a)] at the molecular shell indicate the shell to be the dominant zone of repulsive positron-fullerene interactions; (ii) The shape of the Ps($1s$) radial wavefunction ${\tilde R}_{1s}(\rho)$ as a function of electron-positron separation $\rho=|\vec{r}_+-\vec{r}_-|$ justifies the maximum Ps probability density at $r_-=r_+$; (iii) The radial wave functions of fullerene $i$-th levels of capture [Fig.\,1(a)] ensure that electrons to form Ps are only available at the shell zone. The analytic simplification to interpret the exact numerical results then followed an approximation of the radial integration to obtain the amplitude for a $\pi$ (number of radial node $\eta_r=1$) or a $\sigma$ ($\eta_r=0$) state capture in the following form:
\begin{eqnarray}\label{tot-amp-pi-sig}
T_{\alpha \beta}^-(\vec{k}_i) &\sim& S(\vec{k}_i) - \frac{1}{k_-q}\sin(Qr_c - \ell_t\pi/2 + \eta_r\pi/2) \nonumber \\
             && \times \int dr_- A(r_p)\sin(Q\alpha(r_p)),
\end{eqnarray}
where the momentum transfer vector $\vec{q}=\vec{k}_+ - \vec{k}_i$ and the recoil momentum $Q=k_i-2k_\pm$ for the Ps formation in the forward direction. $S(\vec{k}_i)$ is the contribution of integrations over $1/\rho$ in \eq{pos-amp} which is assumed weak in resonance structures. $r_p$ are the radial positions of dominant contributions representing $\rho = 0$ where Ps($1s$) wavefunctions have their transient maxima at distances $\alpha$ from the molecular radius $r_c$. Obviously, the integral in the above equation spatially dephases the $\sin(Q\alpha)$ modulation, since $\alpha$ varies with $r_-$, retaining maxima in the amplitude only \textit{via} the term $\cos(Qr_c - \ell_t\pi/2+ \eta_r\pi/2)$. The essence of the mechanism is an interference as a function of $Q$: when an odd integer multiple of the half-wavelength of {\em effective} continuum wave as a function of $Q$ fits a distance $r_p$, a rather complicated diffraction pattern in the energy domain is formed from the constructive interference. Several such fringe systems due to the variable $r_p$ cumulatively overlap to finally result into a {\em centroid} fringe pattern of more uniform peaks (bright-spots) \textit{via} a dephasing mechanism in the integration described above. Since these diffraction peaks appear in the energy (momentum) domain, they are characteristically diffraction resonances.
\begin{figure}[h!]
\vskip 0.5cm
\includegraphics[width=8cm]{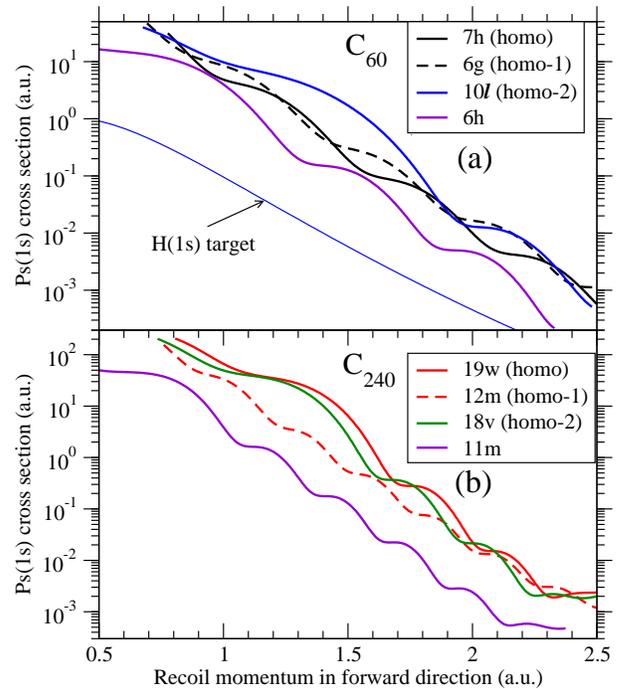}
	\caption{(Color online) (a) Ps($1s$) formation cross sections for captures from four outer levels of $\ful$ (a) and $\full$ (b) as a function of the recoil momentum ($Q$) in forward direction. The corresponding result of hydrogen $1s$ capture is also shown on panel (a) for the comparison.}
	\label{figure3}
\end{figure}

Resonances, however, must scale differently in the cross sections, which are derived from the squared modulus of the amplitude \textit{via} \eq{dcs-nlm}. Therefore, squaring of \eq{tot-amp-pi-sig} results in doubling the argument of the trigonometric function to obtain $\cos(Qd_c - \ell_t\pi + \eta_r\pi)$ for approximate resonance positions in the cross section, where $d_c$ is the fullerene diameter; note that besides $d_c$, the positions also depend on phase-shifts based on initial state information $\ell_t$ and $\eta_r$. In any case, at the cross section level, one may draw an analogy with the single-slit experiment of classical wave optics: Ps formation amplitudes from diametrically opposite sites of a fullerene molecule quantum mechanically interfere to produce fringe patterns in the momentum domain, as schematically shown in Fig.\,2. Consequently, the resonance pattern must shrink for a larger fullerene due to increased slit width as we discuss below.

Fig.\,3 presents the exact numerical Ps($1s$) cross sections as a function of the forward-emission recoil momentum $Q$ that displays series of broad resonances for a set of four capture levels of $\ful$ and $\full$. The range of $Q$ corresponds to the electron excitation energy from roughly 50 eV, which is above the plasmon excitations, to 270 eV, which is below the $K$-shell of atomic carbon (in order to validate the jellium modeling). Note that the Ps($1s$) cross section in Fig.\,3(a) for the capture from the $1s$ level atomic hydrogen is flat, since no diffraction is possible when the electron is captured from ``everywhere'' in a Coulomb system. We also note in Fig.\,3 that, as expected, the non-resonant background strength of the cross sections is proportional to the number of electrons that fills the level, while for a fixed occupancy number (same $\ell_t$) a $\pi$ level produces stronger cross section than a $\sigma$ likely due to larger spatial spread of $\pi$ wavefunctions [Fig.\,1(a)].

Considering the resonances in Fig.\,3, we first note a general trend: the resonances at low $Q$ for the capture from high angular momentum states, such as, homo-2 of $\ful$ and homo, homo-2 of $\full$, are significantly wide. This is likely a direct consequence of stronger distortions of continuum waves for higher $\ell_t$.  For the relative positions of the resonances of varying capture states several observations can be made. As seen in Fig.\,3(a), the resonances for captures from $\ful$ homo ($7h$) versus homo-1 ($6g$) are positioned out-of-phase, since, even though both levels are of $\pi$ character ($\eta_r$ =1), their angular quantum number $\ell_t$ differs by one unit resulting in a 180$^o$ relative phase-shift in $\cos(Qd_c - \ell_t\pi + \eta_r\pi)$. This comparison is however more complicated between homo ($19w$) and homo-1 ($12m$) captures of $\full$ [Fig.\,3(b)], since, not only that these levels are of respectively $\sigma$ and $\pi$ nodal characters, but also their $\ell_t$ values are vastly different. However, an out-of-phase offset between homo and homo-2 ($18v$) resonances of $\full$ is roughly noted at least at higher $Q$, since both are $\sigma$ levels but their $\ell_t$ differs by one. Furthermore, comparing the results between $\ful$ homo and an inner $\sigma$ level $6h$ of the identical angular character, out-of-phase resonance locations are seen -- a pattern which is obviously due to their $\pi$ versus $\sigma$ characters accounting for a half-cycle shift. Likewise, the same reason explains why the homo-1 compared to the inner $11m$ capture in $\full$ displays out-of-phase resonances. All these observations generally indicate that electronic structural information can be accessed spectroscopically by level-differential Ps formation from fullerenes.
\begin{figure}[h!]
\vskip 0.5cm
\includegraphics[width=8cm]{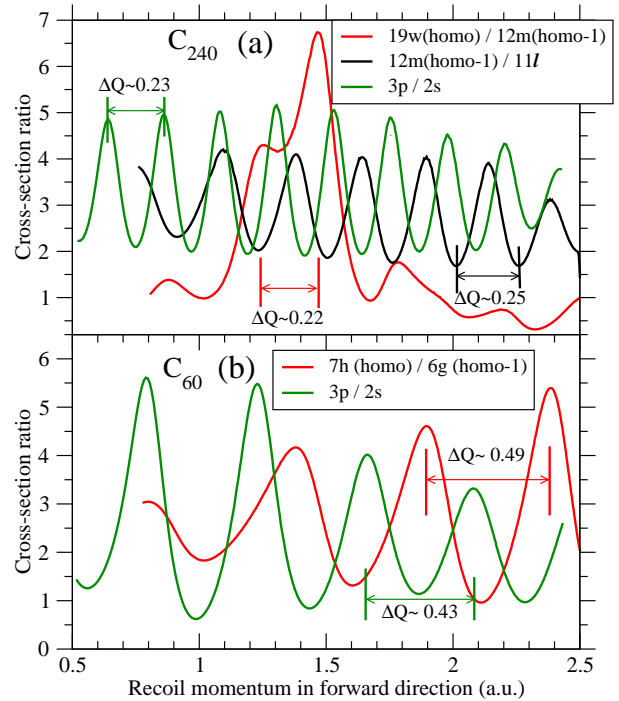}
	\caption{(Color online) Ratios of selected combinations of cross sections, for $\full$ (a) and for $\ful$ (b), illustrate resonances. Typical separations ($\Delta Q$) between some of these resonances are marked.}
	\label{figure4}
\end{figure}

The variations and similarities in the shape of the resonances are most spectacularly illustrated by considering the cross section ratios, shown in Fig.\,4, which neutralize the non-resonant background decays. Accessing these ratios in experiments by the Ps formation spectroscopy may improve the accuracy by minimizing experimental noise from cancellations. It is important to note again that even though we are attempting to use plane wave descriptions in our analysis to interpret the key results, the exact character of the resonances in Fig.\,4 are far more diverse. While the ratios of capture levels of $\pi$ characters, both the combinations chosen for $\ful$ [Fig.\,4(b)] and homo-1/$11l$, $3p$/$2s$ for $\full$ [Fig.\,4(a)], produce reasonably uniform structures, the resonances are dramatically complex for homo/homo-1 of $\full$ where it is a $\pi$-to-$\sigma$ ratio with vastly different angular symmetry. The most noticeable general distinction between the distribution of the resonances of $\ful$ versus $\full$ comes from the molecular size. The underpinning of the diffraction process is evidenced in a nearly halfway shrinkage of the fringe patterns for roughly doubly larger $\full$. This clearly upholds the single-slit analogy. A more quantitative analysis is shown below.

A powerful approach to bring out the connection of diffraction resonances with the fullerene diameter is to evaluate the Fourier spectra of the cross sections as a function of $Q$. To generate the input signals for the Fourier transform of the resonances on a flat, non-decaying background, we considered ratios of the results of two consecutive angular levels of $\pi$ electrons for both the fullerene molecules. Fourier magnitudes of these ratios are calculated by using the fast Fourier transform algorithm after applying an appropriate window function to reduce spurious structures. While such windowing adds some extra width to the ``frequency'' peaks, it practically does not compromise the peak positions. The results are presented in Fig.\,5 in reciprocal (radial) coordinates. All the curves in Fig.\,5(b) exhibit strong peaks located around the diameter $d_c$ for each of $\ful$ and $\full$, as expected from our model equation, \eq{tot-amp-pi-sig}, that includes the function $\cos(Qd_c - \ell_t\pi + \eta_r\pi)$; the transform magnitudes are insensitive to phase-shifts connected to $\ell_t$ and $\eta_r$. To guide the eye, LDA radial potentials and positron scattering potentials for homo captures are plotted in Fig.\,5(a). Notice, the small, systematic offset of the peaks towards lower values with the increasing angular momentum. This is even another signature of the fact that the continuum waves are Coulomb distorted and so are more complicated than simple plane waves used in our model analysis. In fact, we could generally anticipate this variation by noting in Fig.\,4(a) the typical separations $\Delta Q$ of 0.22 a.u.\ , 0.25 a.u.\ and 0.23 a.u.\ respectively for homo/homo-1, homo-1/$11l$ and $3p$/$2s$ ratio and then determining their Fourier conjugate ($2\pi/\Delta Q$) values of 28.5 a.u.\ , 25.1 a.u.\ and 27.3 a.u.\ being somewhat different but close to the radius of 27 a.u.\ of $\full$. But in this case, homo and homo-1, being of $\sigma$ and $\pi$ nodal characters, should be left out in explaining the trend in all-$\pi$ Fourier spectra [Fig.\,5(b)]. For $\ful$, the ratios considered in Fig.\,4(b) are all of $\pi$ symmetry and therefore conform with the Fourier spectra trend. Indeed, the quoted $\Delta Q$ values are 0.49 a.u.\ and 0.43 a.u.\ respectively for homo/homo-1 and $3p$/$2s$ ratio of $\ful$ correspond to slightly increasing Fourier conjugate values of 12.8 a.u.\ and 14.6 a.u., yet being close to the molecular diameter of 13.4 a.u. In summary, these Fourier reciprocal spectra unequivocally support the theme that the host of broad resonances are indeed the fringe patterns in the energy domain for a Ps formation channel where the Ps-emission diffracts in energy off the shell -- a spherical slit. The comparison demonstrates the resonances to be more compact in energy for $\full$, which is a larger diffractor. Our calculations (not shown) for the formation of excited Ps($2s$) have also produced similar general trends.
\begin{figure}[h!]
\includegraphics[width=8.5cm]{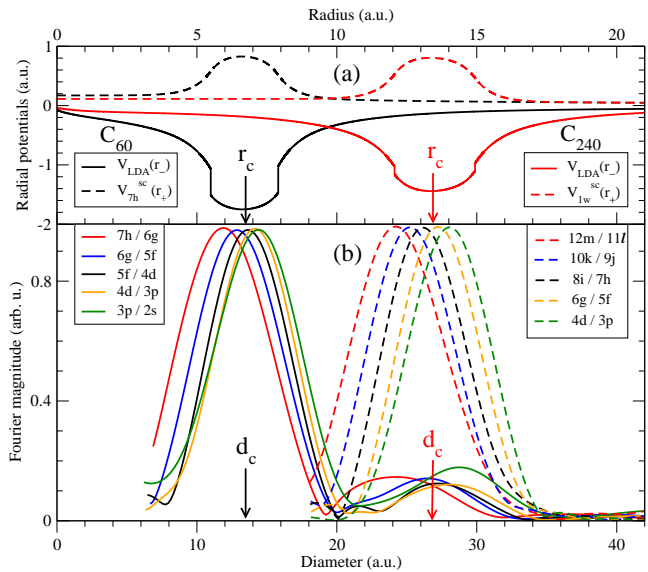}
\caption{(Color online) (a) LDA radial potentials and the positron scattering potentials for the homo capture from both $\ful$ and $\full$; molecular radii are also pointed out. (b) The Fourier transform magnitudes of Ps($1s$) cross section ratios for captures from various $\pi$ levels of both the fullerenes as a function of the radial coordinate$\times$2 with the molecular diameters pointed out.}
\label{figure5}
\end{figure}

\section{Conclusion}

In conclusion, we extend our previous calculation~\cite{hervieux2017} and compare the Ps formation cross sections in the CDW-FS method among electron captures from various electronic levels as well as between $\ful$ and $\full$. The molecular ground state structures are modeled by a simple but successful LDA methodology that used LB94 exchange-correlation functional. Hosts of strong and broad shape resonances in the Ps formation are found that can access electronic structure information of the targets. The resonances engender from a diffraction effect in the Ps formation process localized on the fullerene shell which is further established by comparing results of two different fullerene ``spherical slits". Application of a Fourier analysis technique to the Ps spectra has facilitated the analysis. The success of an analytic model based on forward emissions suggests that the effect is likely predominant in the forward direction of Ps formation.

As an additional future motivation, the Ps($2p$) channel is attractive too, since it can be monitored optically~\cite{murtagh2009}. Also, the effect discussed in this paper should be universal for Ps formation from nanosystems, including metal (alkaline earth, noble, coinage) clusters, carbon nanotubes, or even quantum dots that, like fullerenes, confine finite-sized electron gas. The work further motivates a new research direction to apply Ps formation spectroscopy to gas-phase nanosystems which began with our earlier published research~\cite{hervieux2017,chakraborty2017}, since fullerenes currently enjoy significant attraction in precision measurements. Fullerenes~\cite{thomas2017} and metallic nanoparticles~\cite{jaenkaelae2011,xia2009} are nowadays available in gas-phase. However, probing the target-state differential Ps-signals is still challenging for current techniques~\cite{anderson2014}. But accessing this will be beneficial in general and, in particular, since the predicted resonances, having a target angular-state and radial-structure dependent momentum-shift [\eq{tot-amp-pi-sig}], will largely flatten out in the total Ps measurement. The technique to measure the recoil momentum of the cations may be improved by using a supersonic gas jet to increase the overlap with the positron beam. Resolving the Ps level may not be so critical (may be done by laser spectroscopy of a dense Ps gas), since Ps($1s$) signal should largely dominate. We hope that this theoretical effort will help add further to the motivation in measuring differential Ps production at least within a narrow forward angle.

\begin{acknowledgments}
The research is supported by the National Science Foundation Grant No.\ PHY-1806206, USA.
\end{acknowledgments}


\end{document}